\documentclass[12pt,draftcls,onecolumn]{IEEEtran}
\usepackage{cite}
\usepackage[font=small]{caption}
\usepackage{ifpdf}
\usepackage{authblk}
\usepackage{commath}
\usepackage{cuted} 

\usepackage{bigints}
\newif\ifCLASSOPTIONonecolumn       \CLASSOPTIONonecolumnfalse
\newif\ifCLASSOPTIONtwocolumn       \CLASSOPTIONtwocolumntrue
\ifCLASSINFOpdf
   \usepackage[pdftex]{graphicx}
 \else
    \usepackage[dvips]{graphicx}
  \fi
\usepackage{subcaption}
\usepackage{amsmath}
\usepackage{dsfont}
\usepackage{bbold}
\usepackage{amssymb}
\usepackage{algorithmic}
\usepackage{array}
\usepackage{stix}
\usepackage{stfloats}
\fnbelowfloat
\usepackage{bigints}
\begin{document}
%
\title{Outage Analysis of Cooperative NOMA in Millimeter Wave Vehicular Network at Intersections}
%
%
%
\author[1 2]{Baha Eddine Youcef~Belmekki}
\author[1]{Abdelkrim ~Hamza}
\author[2]{Beno\^it~Escrig}
\affil[1]{LISIC Laboratory, Electronic and Computer Faculty, USTHB, Algiers, Algeria,}
\affil[ ]{email: $\{$bbelmekki, ahamza$\}$@usthb.dz}
\affil[2]{University of Toulouse, IRIT Laboratory, School
of ENSEEIHT, Institut National Polytechnique de Toulouse, France, e-mail: $\{$bahaeddine.belmekki, benoit.escrig$\}$@enseeiht.fr}
\affil[ ]{}
\setcounter{Maxaffil}{0}
\renewcommand\Affilfont{\small}
\markboth{ }%
{Shell \MakeLowercase{\textit{et al.}}: Bare Demo of IEEEtran.cls for IEEE Journals}
\maketitle

\IEEEpeerreviewmaketitle
\begin{abstract}
In this paper, we study cooperative non-orthogonal multiple access scheme (NOMA) for millimeter wave (mmWave) vehicular networks at intersection roads. The intersection consists of two perpendicular roads. Transmissions occur between a source, and two destinations nodes with a help of a relay. We assume that the interference come from a set of vehicles that are distributed as a one dimensional homogeneous Poisson point process (PPP). Our analysis includes the effects of blockage from buildings at intersections. We derive closed form outage probability expressions for cooperative NOMA, and compare them with cooperative orthogonal multiple access (OMA). We show that cooperative NOMA offers a significant improvement over cooperative OMA, especially for high data rates. We also show that as the nodes reach the intersection, the outage probability increases. Counter-intuitively, we show that the non line of sigh (NLOS) scenario has a better performance than the line of sigh (LOS) scenario. The analysis is verified with Monte Carlo simulations.  
\end{abstract}

\subsection{Motivation}
Road traffic safety is a major issue, and more particularly at intersections \cite{traficsafety}. 
Vehicular communications provide helpful applications for road safety and traffic management. These applications help to prevent accidents or alerting vehicles of accidents happening in their surroundings. Hence, these applications require high bandwidth and high spectral efficiency, to insure high reliability and low latency communications.
In this context, non-orthogonal multiple access (NOMA) has been show to increase the data rate and spectral efficiency \cite{ding2017application}. Unlike orthogonal multiple access (OMA), NOMA allows multiple users to share the same resource with different power allocation levels.
On the other hand, the needs of vehicular communications for the fifth generation (5G) in terms of resources require a larger bandwidth. Since the spectral efficiency of sub-6 GHz bands has already reached the theoretical limits, millimeter wave (mmWave) frequency bands (20-100 GHz and beyond) offer a very large bandwidth \cite{roh2014millimeter}.


\subsection{Related Works}
\subsubsection{Cooperative NOMA}
NOMA is an efficient multiple access technique for spectrum use. It has been shown that NOMA outperforms OMA \cite{mobini2017full}.  However, few research investigates the effect of co-channel interference and their impact on the performance considering direct transmissions \cite{ali2018analyzing,zhang2016stochastic,tabassum2017modeling}, and cooperative transmissions \cite{liu2017non,ding2016relay}.
\subsubsection{Cooperative mmWave}
In mmWave bands, few works studied cooperative communications using tools from stochastic geometry \cite{biswas2016performance,wu2017coverage,belbase2018two,belbase2018coverage}. However, in \cite{biswas2016performance,wu2017coverage,belbase2018two}, the effect of small-scale fading is not taken into consideration. In \cite{belbase2018coverage}, the authors investigate the performance of mmWave relaying networks in terms of coverage probability with best relay selection.

\subsubsection{Vehicular communications at intersections}
Several works studied the effect of the interference at intersections, considering OMA. The performance in terms of success probability are derivated considering direct transmission in \cite{steinmetz2015stochastic,abdulla2016vehicle}. The performance of vehicle to vehicle (V2V) communications are evaluated for multiple intersections schemes considering direct transmission in \cite{jeyaraj2017reliability}. In \cite{kimura2017theoretical}, the authors derive the outage probability of a V2V communications with power control strategy of a direct transmission. 
In \cite{article}, the authors investigate the impact of a line of sight and non line of sight transmissions at intersections considering Nakagami-$m$ fading channels. The authors in \cite{belmekki2018performance} study the effect of mobility of vehicular communications at road junctions.
In \cite{WCNC,VTC,WiMob,J3,J4}, the authors respectively study the impact of non-orthogonal multiple access, cooperative non-orthogonal multiple access, and maximum ratio combining with NOMA at intersections.
Following this line of research, we study the performance of VCs at intersections in the presence of interference. 

Following this line of research, we study the performance of vehicular communications at intersections in the presence of interference. 
In this paper, the authors extend their work \cite{NoMa} to cooperative transmissions using NOMA  considering mmWave networks. Our analysis includes the effects of blockage from the building in intersections, and Nakagami-$m$ fading channels between the transmitting nodes with difference values of $m$ for LOS and NLOS are considered. Unlike other works that uses approximations, closed form expressions are obtained for Nakagami-$m$ fading channel.

\subsection{Contributions}
The  contributions of this paper are  as follows:
\begin{itemize}
\item
We study the impact and the improvement of using cooperative NOMA on a mmWave vehicular network at intersection roads. Closed form expressions of the outage probability are obtained.
\item
Our analysis includes the effects of blockage from the building in intersections, and Nakagami-$m$ fading channels with difference values of $m$ for LOS and NLOS are considered.

\item
We evaluate the performance of NOMA for both intersection, and show that the outage probability increases when the vehicles move toward the intersections. We also show the effect of LOS and NLOS on the performance at the intersection.

\item We compare all the results obtained with cooperative OMA, and show that cooperative NOMA is superior in terms of outage probability than OMA.
\end{itemize}

\section{System Model}

\begin{figure}[]
\centering
\includegraphics[scale=0.7]{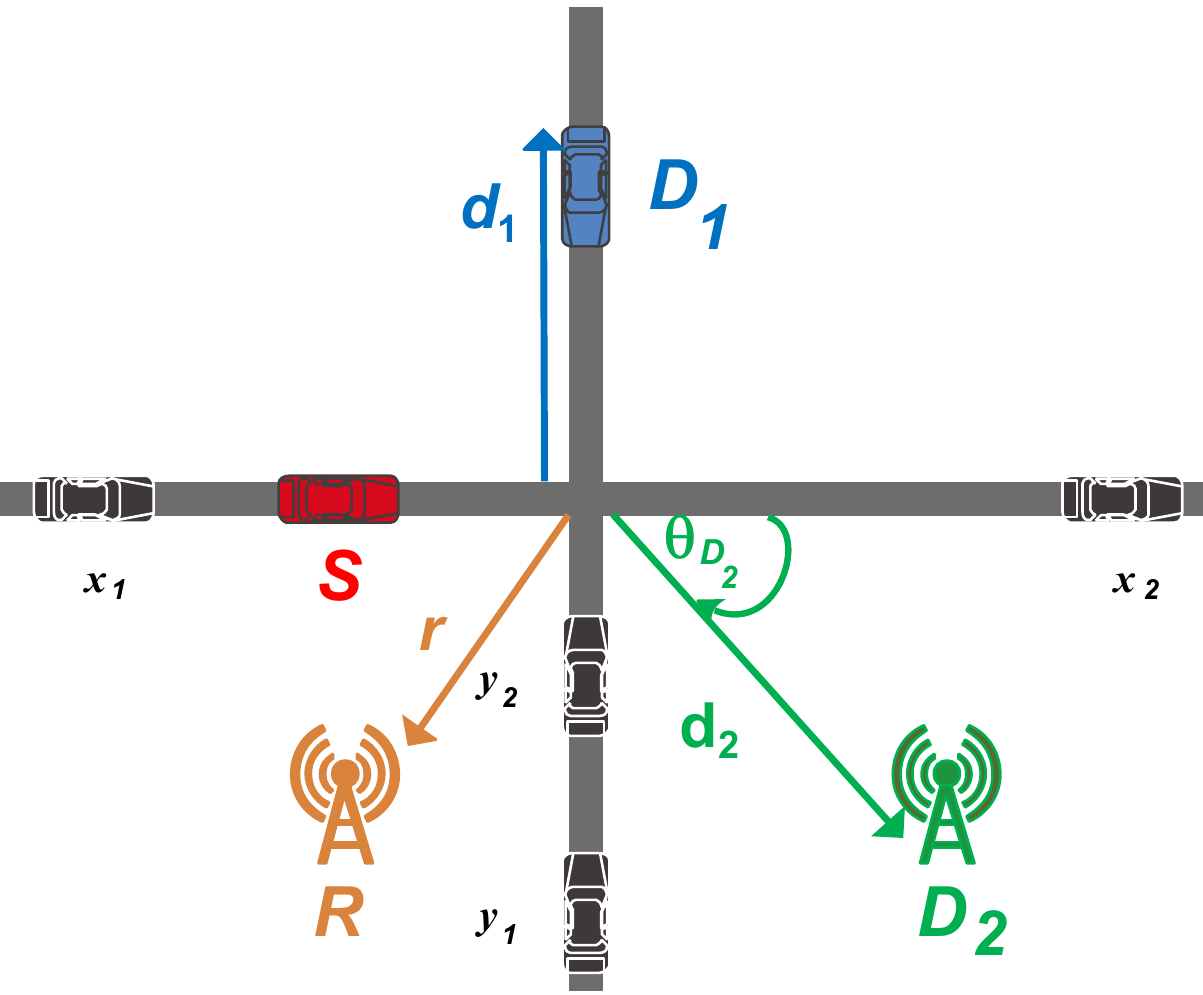}
\caption{Cooperative NOMA system model for vehicular communications involving one relay two receiving node. The receiving nodes can be vehicles or as part of the communication infrastructure. For instance, $S$ and $D_1$ are vehicles, and $R$ and $D_2$ are infrastructures.}
\label{Fig1}
\end{figure}
\subsection{Scenario Model}
In this paper, we consider a mm-Wave vehicular network using a cooperative NOMA transmission between a source, denoted $S$, and two destinations denoted $D_1$ and $D_2$ with the help of a relay denoted $R$. The set $\{S, R, D_1, D_2\}$ denotes the nodes and their locations as depicted in Fig.1.

We consider, an intersection scenario involving two perpendicular roads, an horizontal road denoted by $X$, and a vertical road denoted by $Y$.
In this paper, we consider both V2V and V2I communications\footnote{The Doppler shift and time-varying effect of V2V and V2I channels is beyond the scope of this paper.}, hence, any node of the set $\lbrace{S, R, D_1, D_2}\rbrace$ can be on the road or outside the roads. We denote by  $M$ the receiving node, and by $m$ the distance between the node $M$ and the intersection, where $M \in \{R,D_1,D_2\}$ and $m \in \{r,d_1,d_2\}$, as shown in Fig.1. The angle $\theta_{M} $ is the angle between the node ${M}$ and the X road (see Fig.1). Note that the intersection is the point where the $X$ road and the $Y$ road intersect.
The set $\lbrace{S, R, D_1, D_2}\rbrace$ is subject to interference that are originated from vehicles located on the roads. 

The set of interfering vehicles located on the $X$ road that are in a LOS with $\lbrace{S, R, D_1, D_2}\rbrace$, denoted by $\Phi^{\textrm{LOS}}_{X}$ (resp. on axis $Y$, denoted by $\Phi^{\textrm{LOS}}_{Y}$) are modeled as a One-Dimensional Homogeneous Poisson Point Process (1D-HPPP), that is, $\Phi^{\textrm{LOS}}_{X}\sim\textrm{1D-HPPP}(\lambda^{\textrm{LOS}}_{X},x)$ (resp.$\Phi^{\textrm{LOS}}_{Y}$ $\sim \textrm{1D-HPPP}(\lambda^{\textrm{LOS}}_{Y},y)$, where $x$ and $\lambda^{\textrm{LOS}}_{X}$ (resp. $y$ and $\lambda^{\textrm{LOS}}_{Y}$) are the position of the LOS interferer vehicles and their intensity on the $X$ road (resp. $Y$ road). 

Similarly, the set of interfering vehicles located on the $X$ road that are in a NLOS with $\lbrace{S, R, D_1, D_2}\rbrace$, denoted by $\Phi^{\textrm{NLOS}}_{X}$ (resp. on axis $Y$, denoted by $\Phi^{\textrm{NLOS}}_{Y}$) are modeled as a One-Dimensional Homogeneous Poisson Point Process (1D-HPPP), that is, $\Phi^{\textrm{NLOS}}_{X}\sim\textrm{1D-HPPP}(\lambda^{\textrm{NLOS}}_{X},x)$ (resp.$\Phi^{\textrm{NLOS}}_{Y}$ $\sim \textrm{1D-HPPP}(\lambda^{\textrm{NLOS}}_{Y},y)$, where $x$ and $\lambda^{\textrm{NLOS}}_{X}$ (resp. $y$ and $\lambda^{\textrm{NLOS}}_{Y}$) are the position of the NLOS interferer vehicles and their intensity on the $X$ road (resp. $Y$ road). The notation $x$ and $y$ denotes both the interferer vehicles and their locations. 
\subsection{Blockage Model}
At the intersection, the mmWave signals cannot penetrate the buildings and other obstacles, which causes the link to be in LOS, or in NLOS. 
The event of a link between a node $a$ and $b$ is in a LOS and NLOS, are respectively defined as $\textrm{LOS}_{ab}$, and $\textrm{NLOS}_{ab}$.
The LOS probability function $\mathbb{P}(\textrm{LOS}_{ab})$ is used, where the link between $a$ and $b$
has a LOS probability $\mathbb{P}(\textrm{LOS}_{ab}) = \exp(-\beta r_{ab} )$ and NLOS probability
$\mathbb{P}(\textrm{NLOS}_{ab}) = 1-\mathbb{P}(\textrm{LOS}_{ab})$, where the constant rate $\beta$
depends on the building size, shape and density \cite{bai2014analysis}.

\subsection{Transmission and Decoding Model}
The transmission is subject to a path loss, denoted by $r_{ab}^{-\alpha}$ between the nodes $a$ and $b$, where $ r_{ab}=\Vert a- b\Vert$, and $\alpha$ is the path loss exponent. The path exponent $\alpha \in \{\alpha_{\textrm{LOS}}, \alpha_{\textrm{NLOS}}\}$, where $\alpha=\alpha_{\textrm{LOS}}$, when the transmission is in LOS, whereas $\alpha=\alpha_{\textrm{NLOS}}$, when transmission is in NLOS.

We consider slotted ALOHA protocol with parameter $p$, i.e., every node accesses the medium with a probability $p$. 

 We use a Decode and Forward (DF) decoding strategy, i.e., $R$ decodes the message, re-encodes it, then forwards it to $D_1$ and $D_2$. We also use a half-duplex transmission in which a transmission occurs during two phases. Each phase lasts one time slot. During the first phase, $S$ broadcasts the message to $R$ ($S \rightarrow R$). During the second phase, $R$ broadcasts the message to $D_1$ and $D_2$ ($R \rightarrow D_1$ and $R \rightarrow D_2$).
\subsection{NOMA Model}
We consider, in this paper, that the receiving nodes, $D_1$ and $D_2$, are ordered according to their quality of service (QoS) priorities \cite{ding2016relay,ding2016mimo}. We consider the case when node $D_1$ needs a low data rate but has to be served immediately, whereas node $D_2$ requires a higher data rate but can be served later. For instance, $D_1$ can be a vehicle that needs to receive safety data information about an accident in its surrounding, whereas $D_2$ can be a user that accesses the internet connection.
\subsection{Directional Beamforming Model}

We model the directivity similar to in \cite{singh2015tractable}, where the directional gain, denoted $G(\omega)$, within
the half power beamwidth ($\phi/2$) is $G_{max}$ and is $G_{min}$ in all other
directions. The gain is then expressed as
\begin{equation}
   G(\omega)=\left\{
                \begin{array}{ll}
                  G_{max}, &  \textrm{if} \:\:|\omega|\leq \frac{\phi}{2};\\
                  G_{min}, & \textrm{otherwise}.      
                \end{array}
              \right.
\end{equation}
In this paper, we consider a perfect beam alignment between the nodes, hence $G_{eq} = G_{max}^2$. The impact of beam misalignment is beyond the scope of this paper. 

\subsection{Channel and Interference Model}

 We consider an interference limited scenario, that is, the power of noise is set to zero ($\sigma^2=0$). Without loss of generality, we assume  that all nodes transmit with a unit power. 
The signal transmitted by $S$, denoted $ \chi_{S}$ is a mixture of the message intended to $D_1$ and $D_2$. This can be expressed as
\begin{equation}
 \chi_{S}=\sqrt{a_1}\chi_{D1}+\sqrt{a_2}\chi_{D2}, \nonumber
 \end{equation}
where $a_i$ is the power coefficients allocated to $D_i$, and $\chi_{Di}$ is the message intended to $D_i$, where $i \in \{1,2\}$. Since $D_1$ has higher power than $D_2$, that is $a_1 \ge a_2$, then $D_1$ comes first in the decoding order. Note that, $a_1+a_2=1$.\\
 The signal received at $R$ during the first time slot is expressed as
 
 \begin{align}
   \mathcal{Y}_{R}=&h_{SR}\sqrt{ r_{SR}^{-\alpha_{\textrm{LOS}}}\Upsilon}\:\chi_{S}\mathds{1}(\textrm{LOS}_{SR})
   +h_{SR}\sqrt{ r_{SR}^{-\alpha_{\textrm{NLOS}}}\Upsilon}\:\chi_{S}\mathds{1}(\textrm{NLOS}_{SR}) \nonumber\\
 &+\sum_{x\in \Phi^{\textrm{LOS}}_{X_{R}}}h_{Rx}\sqrt{r_{Rx}^{-\alpha_{\textrm{LOS}}}\Upsilon}\:\chi_{x} 
 +\sum_{y\in \Phi^{\textrm{LOS}}_{Y_{R}}}h_{Ry}\sqrt{r_{Ry}^{-\alpha_{\textrm{LOS}}}\Upsilon}\:\chi_{y}\nonumber\\
 &+\sum_{x\in \Phi^{\textrm{NLOS}}_{X_{R}}}h_{Rx}\sqrt{r_{Rx}^{-\alpha_{\textrm{NLOS}}}\Upsilon}\:\chi_{x} 
 +\sum_{y\in \Phi^{\textrm{NLOS}}_{Y_{R}}}h_{Ry}\sqrt{r_{Ry}^{-\alpha_{\textrm{NLOS}}}\Upsilon}\:\chi_{y}. \nonumber
 \end{align}
 The signal received at $D_i$ during the second time slot is expressed as
 \begin{align}
   \mathcal{Y}_{D_i}=&h_{RD_i}\sqrt{r_{RD_i}^{-\alpha}\Upsilon}\:\chi_{R}\mathds{1}(\textrm{LOS}_{RD_i})
   + h_{RD_i}\sqrt{r_{RD_i}^{-\alpha}\Upsilon}\:\chi_{R}\mathds{1}(\textrm{NLOS}_{RD_i})\nonumber\\
&+ \sum_{x\in \Phi^{\textrm{LOS}}_{X_{D_i}}}h_{Dix}\sqrt{r_{D_{i}x}^{-\alpha_{\textrm{LOS}}}\Upsilon}\:\chi_{x} 
 +\sum_{y\in \Phi^{\textrm{LOS}}_{Y_{D_i}}}h_{Diy}\sqrt{r_{D_{i}y}^{-\alpha_{\textrm{LOS}}}\Upsilon}\:\chi_{y}\nonumber\\
 &+\sum_{x\in \Phi^{\textrm{NLOS}}_{X_{D_i}}}h_{Dix}\sqrt{r_{D_{i}x}^{-\alpha_{\textrm{NLOS}}}\Upsilon}\:\chi_{x} 
 +\sum_{y\in \Phi^{\textrm{NLOS}}_{Y_{D_i}}}h_{Diy}\sqrt{r_{D_{i}y}^{-\alpha_{\textrm{NLOS}}}\Upsilon}\:\chi_{y}, \nonumber
 \end{align}
where $\mathcal{Y}_{M}$ is the signal received by $M$, and $\chi_{R}$ is the message transmitted by $R$.
The messages transmitted by the interfere node $x$ and $y$, are denoted respectively by $ \chi_x$ and $\chi_y $. The term $\Upsilon= G_{eq} \eta^2/ (4 \pi)^2$ models the directional gain, the reference path loss at one meter, and $\eta$ is the wavelength of the operating frequency.

The coefficients $h_{SR}$, and $h_{RD_i}$ denote the fading of the link $S-R$, and $R-D_i$. The fading coefficients are distributed according to a Nakagami-$m$ distribution with parameter $m$ \cite{belbase2018coverage}, that is 
\begin{equation}
f_{h_u}(x)=2 \Big(\frac{m}{\mu}\Big)^m \frac{x^{2m-1}}{\Gamma(m)}e^{-\frac{m}{\mu}x^2},
\end{equation}
where $u\in \{SR,RD_i\}$. The parameter $m \in \{m_{\textrm{LOS}}, m_{\textrm{NLOS}}\}$, where $m=m_{\textrm{LOS}}$ when $u$ is in a LOS, whereas $m=m_{\textrm{NLOS}}$, when $u$ is in a NLOS. The parameter $\mu$ is the average
received power.

Hence, the power fading coefficients $|h_{SR}|^2$, and $|h_{RD_i}|^2$ are distributed according to a gamma distribution, that is, 
\begin{equation}
f_{| h_u | ^2}(x)=\Big(\frac{m}{\mu}\Big)^m \frac{x^{m-1}}{\Gamma(m)}e^{-\frac{m}{\mu}x}.
\end{equation}

The fading coefficients $h_{Rx}$,$h_{Ry}$,$h_{D_{i}x}$ and  $h_{D_iy}$ denote the fading of the link $R-x$, $R-y$, $D_i-x$, and $D_i-y$. The fading coefficients are modeled as Rayleigh fading \cite{deng2017meta}. Thus, the power fading coefficients $|h_{Rx}|^2$, $|h_{Ry}|^2$ $|h_{D_ix}|^2$ and $|h_{D_iy}|^2$, are distributed according to an exponential distribution with unit mean.

The aggregate interference is defined as from the $X$ road at $M$, denoted $I_{X_{M}}$, is expressed as
\begin{equation}\label{EQ.1}
I_{X_{M}}= I^{\textrm{LOS}}_{X_{M}} +I^{\textrm{NLOS}}_{X_{M}}   
=\sum_{x\in \Phi^{\textrm{LOS}}_{X_{M}}}\vert h_{Mx}\vert^{2}r_{Mx}^{-\alpha_{\textrm{LOS}}}\Upsilon + \sum_{y\in \Phi^{\textrm{NLOS}}_{X_{M}}}\vert h_{Mx}\vert^{2}r_{Mx}^{-\alpha_{\textrm{NLOS}}}\Upsilon, 
\end{equation}
where $I^{\textrm{LOS}}_{X_{M}} $ denotes the aggregate interference from the $X$ road that are in a LOS with $M$, and $I^{\textrm{NLOS}}_{X_{M}} $ denotes the aggregate interference from the $X$ road that are in a NLOS with $M$. 
Similarly, $\Phi^{\textrm{LOS}}_{X_{M}}$ and $\Phi^{\textrm{NLOS}}_{X_{M}}$, denote respectively, the set of the interferers from the $X$ road at $M$ in a LOS, and in NLOS.

In the same way, the aggregate interference is defined as from the $Y$ road at $M$, denoted $I_{Y_{M}}$, is expressed as
\begin{equation}\label{EQ.2}
I_{Y_{M}}=  I^{\textrm{LOS}}_{Y_{M}} + I^{\textrm{NLOS}}_{Y_{M}} 
= \sum_{y\in \Phi^{\textrm{LOS}}_{Y_{M}}}\vert h_{My}\vert^{2}r_{My}^{-\alpha_{\textrm{LOS}}}\Upsilon +\sum_{y\in \Phi^{\textrm{NLOS}}_{Y_{M}}}\vert h_{My}\vert^{2}r_{My}^{-\alpha_{\textrm{NLOS}}}\Upsilon, 
\end{equation}
where $I^{\textrm{LOS}}_{Y_{M}} $ denotes the aggregate interference from the $X$ road that are in a LOS with $M$, and $I^{\textrm{NLOS}}_{Y_{M}} $ denotes the aggregate interference from the $Y$ road that are in a NLOS with $M$. 
Similarly, $\Phi^{\textrm{LOS}}_{Y_{M}}$ and $\Phi^{\textrm{NLOS}}_{Y_{M}}$, denote respectively, the set of the interferers from the $Y$ road at $M$ in a LOS, and in NLOS.

\section{Cooperative NOMA Outage Expressions}
\subsection{Signal-to-Interference Ratio (SIR) Expressions}
We define the outage probability as the probability that the signal-to-interference ratio (SIR) at the receiver is below a given threshold. According to successive interference cancellation (SIC) \cite{hasna2003performance}, $D_1$ will be decoded first at the receiver since it has the higher power allocation, and $D_2$ message will be considered as interference. The SIR at $R$ to decode $D_1$, denoted  $\textrm{SIR}^{(\alpha)}_{R_1}$, is expressed as
\begin{equation}\label{EQ.3}
\textrm{SIR}^{(\alpha)}_{R_1}=\frac{\vert h_{SR}\vert^{2}r_{SR}^{-\alpha} \Upsilon \,a_1}{\vert h_{SR}\vert^{2}r_{SR}^{-\alpha}\Upsilon a_2+I_{X_{R}}+I_{Y_{R}}} .
\end{equation}
Since $D_2$ has a lower power allocation, $R$ has to decode $D_1$ message, then decode $D_2$ message. The SIR at $R$ to decode $D_2$ message, denoted $\textrm{SIR}^{(\alpha)}_{R_2}$, is expressed as \footnote{Perfect SIC is considered in this work, that is, no fraction of power remains after the SIC process.}

\begin{equation}\label{EQ.4}
\textrm{SIR}^{(\alpha)}_{R_2}=\frac{\vert h_{SR}\vert^{2}r_{SR}^{-\alpha}\Upsilon \,a_2}{I_{X_{R}}+I_{Y_{R}}}.
\end{equation}
The SIR at $D_1$ to decode its intended message, denoted $\textrm{SIR}^{(\alpha)}_{D_1}$, is given by

\begin{equation}\label{EQ.5}
\textrm{SIR}^{(\alpha)}_{D_1}=\frac{\vert h_{RD1}\vert^{2}r_{RD1}^{-\alpha}\Upsilon \,a_1}{\vert h_{RD1}\vert^{2}r_{RD1}^{-\alpha} \Upsilon a_2+I_{X_{D1}}+I_{Y_{D1}}} .
\end{equation}
In order for $D_2$ to decode its intended message, it has to decode $D_1$ message. The SIR at $D_2$ to decode $D_1$ message, denoted $\textrm{SIR}^{(\alpha)}_{D_{2-1}}$, is expressed as
\begin{equation}\label{EQ.6}
\textrm{SIR}^{(\alpha)}_{D_{2-1}}=\frac{\vert h_{RD2}\vert^{2}r_{RD2}^{-\alpha}\Upsilon \,a_1}{\vert h_{RD2}\vert^{2}r_{RD2}^{-\alpha}\Upsilon a_2+I_{X_{D2}}+I_{Y_{D2}}}.
\end{equation}
The SIR at $D_2$ to decode its intended message, denoted $\textrm{SIR}^{(\alpha)}_{D_{2}}$, is expressed as 
\begin{equation}\label{EQ.7}
\textrm{SIR}^{(\alpha)}_{D_2}=\frac{\vert h_{RD2}\vert^{2}r_{RD2}^{-\alpha}\Upsilon \,a_2}{I_{X_{D2}}+I_{Y_{D2}}}.
\end{equation}
\subsection{Outage Event Expressions}
The outage event that $R$ does not decode $D_1$ message, denoted $\textit{O}_{R_1}$, is given by
\begin{equation}\label{EQ.8}
\textit{O}_{R_1}\triangleq \bigcup_{{\textrm{Z}}\in \{\textrm{LOS},\textrm{NLOS}\}} \big\{ \textrm{Z}_{SR}\cap (\textrm{SIR}^{(\alpha_{\textrm{Z}})}_{R_1}< \Theta_1) \big\} ,  
\end{equation}
where $\Theta_1=2^{2\mathcal{R}_1}-1$, and $\mathcal{R}_1$ is the target data rate of $D_1$.\\
Also, the outage event that $D_1$ does not decode its intended message, denoted $\textit{O}_{D_1}$, is given by
\begin{equation}\label{EQ.9}
\textit{O}_{D_1}\triangleq \bigcup_{{\textrm{Z}}\in \{\textrm{LOS},\textrm{NLOS}\}} \big\{ \textrm{Z}_{RD_1} \cap (\textrm{SIR}^{(\alpha_{\textrm{Z}})}_{D_1}< \Theta_1) \big\} ,  
\end{equation}
Then, the overall outage event related to $D_1$, denoted $\textit{O}_{(1)}$, is given by 
\begin{equation}\label{EQ.10}
\textit{O}_{(1)}\triangleq  \left[\textit{O}_{R_1} \cup \textit{O}_{D_1}\right],  
\end{equation}
The outage event that $R$ does not decode $D_2$ message, denoted $\textit{O}_{R_2}$, is given by
\begin{equation} \label{EQ.11}
\textit{O}_{R_2}\triangleq \bigcup_{{\textrm{Z}}\in \{\textrm{LOS},\textrm{NLOS}\}}\bigcup_{i=1}^{2} \big\{ \textrm{Z}_{SR} \cap (\textrm{SIR}^{(\alpha_{\textrm{Z}})}_{R_i}< \Theta_i) \big\} ,  
\end{equation}
where $\Theta_2=2^{2\mathcal{R}_2}-1$ ($i=2)$, and $\mathcal{R}_2$ is the target data rate of $D_2$.
Also, the outage event that $D_2$ does not decode its intended message, denoted $\textit{O}_{D_2}$, is given by
\begin{equation}\label{EQ.12}
\textit{O}_{D_2}\triangleq \bigcup_{{\textrm{Z}}\in \{\textrm{LOS},\textrm{NLOS}\}}\bigcup_{i=1}^{2} \big\{ \textrm{Z}_{RD_2} \cap (\textrm{SIR}^{(\alpha_{\textrm{Z}})}_{D_{2-i}}< \Theta_i) \big\} ,  
\end{equation}
Finally, the overall outage event related to $D_2$, denoted $\textit{O}_{(2)}$, is given by 
\begin{equation}\label{EQ.13}
\textit{O}_{(2)}\triangleq \left[\textit{O}_{R_2} \cup \textit{O}_{D_2}\right] .  
\end{equation}
\subsection{Outage Probability Expressions}
In the following, we will express the outage probability related to  $\textit{O}_{(1)}$ and $\textit{O}_{(2)}$.  The probability  $\mathbb{P}(\textit{O}_{(1)})$ is given, when $\Theta_1 < \frac{a_1}{a_2}$, by  (\ref{EQ.14}) 
\begin{equation}\label{EQ.14}
\mathbb{P}(\textit{O}_{(1)})=1-
\Bigg\{ \sum_{\textrm{Z}\in\{\textrm{LOS},\textrm{NLOS}\}}^{} \mathbb{P}(\textrm{Z}_{SR}) \Lambda\Big(\dfrac{m_\textrm{Z}\:\Psi_1}{\mu \:r_{SR}^{-\alpha_\textrm{Z}}\Upsilon }\Big) \times \sum_{\textrm{Z}\in\{\textrm{LOS},\textrm{NLOS}\}}^{} \mathbb{P}(\textrm{Z}_{RD_1}) \Lambda\Big(\dfrac{m_\textrm{Z}\:\Psi_1}{\mu \:r_{RD_1}^{-\alpha_\textrm{Z}}\Upsilon }\Big)  \Bigg\},
\end{equation}
where $\Psi_{1}=\Theta_1/(a_1-\Theta_1a_2)$. The expression of $\Lambda\Big(\dfrac{m\:\Psi}{\mu \:r_{ab}^{-\alpha}\Upsilon } \Big)$ is given by
\begin{multline}\label{EQ.16}
\Lambda\Big(\dfrac{m\:\Psi}{\mu \:r_{ab}^{-\alpha}\Upsilon } \Big)=\\ \prod_{\textrm{K}\in\{\textrm{LOS},\textrm{NLOS}\}} \sum_{k=0}^{m-1}\frac{1}{k!}\:\big(-\dfrac{m\:\Psi}{\mu \:r_{ab}^{-\alpha} \Upsilon}\big)^k \sum_{n=0}^{k}\binom{k}{n}\frac{\textrm{d}^{k-n} \mathcal{L}_{I^{\textrm{K}}_{X_b}}\big(\dfrac{m\:\Psi}{\mu \:r_{ab}^{-\alpha}\Upsilon }\big)}{\textrm{d}^{k-n} \big(\dfrac{m\:\Psi}{\mu \:r_{ab}^{-\alpha_L}\Upsilon }\big)} \frac{\textrm{d}^{n} \mathcal{L}_{I^{\textrm{K}}_{Y_b}}\big(\dfrac{m\:\Psi}{\mu \:r_{ab}^{-\alpha}\Upsilon }\big)}{\textrm{d}^{n} \big(\dfrac{m\:\Psi}{\mu \:r_{ab}^{-\alpha} \Upsilon}\big)}.
\end{multline}
The  probability $ \mathbb{P}(\textit{O}_{(2)})$ is given, when $\Theta_1 < \frac{a_1}{a_2}$, by (\ref{EQ.15})
\begin{equation}\label{EQ.15}
\mathbb{P}(\textit{O}_{(2)})=1-
\Bigg\{ \sum_{\textrm{Z}\in\{\textrm{LOS},\textrm{NLOS}\}}^{} \mathbb{P}(\textrm{Z}_{SR}) \Lambda\Big(\dfrac{m_\textrm{Z}\:\Psi_{\mathrm{max}}}{\mu \:r_{SR}^{-\alpha_\textrm{Z}}\Upsilon }\Big) \times \sum_{\textrm{Z}\in\{\textrm{LOS},\textrm{NLOS}\}}^{} \mathbb{P}(\textrm{Z}_{RD_2}) \Lambda\Big(\dfrac{m_\textrm{Z}\:\Psi_{\mathrm{max}}}{\mu \:r_{RD_2}^{-\alpha_\textrm{Z}}\Upsilon }\Big)  \Bigg\},
\end{equation}
where $\Psi_{\mathrm{max}}=\mathrm{max}(\Psi_1,\Psi_2)$, and $\Psi_2=\Theta_2/a_2$.\\
\textit{Proof}:  See Appendix A.\hfill $ \blacksquare $ \\
\begin{figure}[]
\centering
\includegraphics[height=8cm,width=9cm]{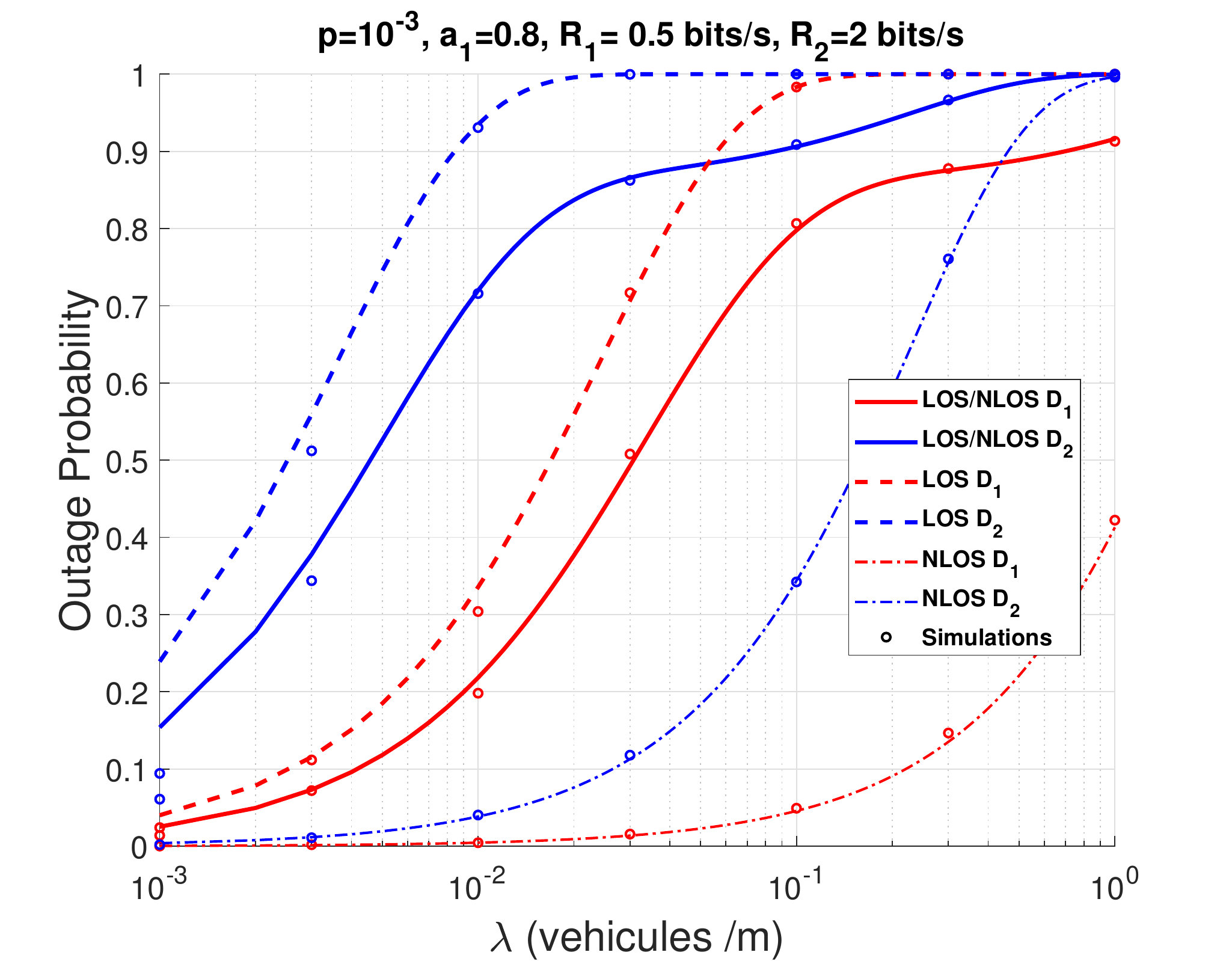}
\caption{Outage probability as function of $\lambda$ considering cooperative  NOMA, for LOS transmission, NLOS, and LOS/NLOS (the equation (\ref{EQ.14}) and (\ref{EQ.15})).}
\label{Fig3}
\end{figure}
 \section{Laplace Transform Expressions}
We present the Laplace transform expressions of the interference from the X road at the receiving node denoted by $M$, denoted $\mathcal{L}_{I^{\textrm{K}}_{X_{M}}}$, and from the Y road at the receiving node denoted by $M$, denoted $\mathcal{L}_{I^{\textrm{K}}_{Y_{M}}}$. We only present the case when $\alpha_{\textrm{K}}=2$ due to the lack of space. The Laplace transform expressions of the interference at the node $M$  for an intersection scenario, when  $\alpha_{\textrm{K}}=2$ are given by 
\begin{equation}\label{EQ.17}
\mathcal{L}_{I^{\textrm{K}}_{X_{M}}}(s)=\exp\Bigg(\dfrac{-\emph{p}\lambda^{\textrm{K}}_{X}s\pi}{\sqrt{\big[m\sin(\theta_{{M}})\big]^2+s}}\Bigg),
\end{equation}
and
\begin{equation}\label{EQ.18}
\mathcal{L}_{I^{\textrm{K}}_{Y_{M}}}(s)=\exp\Bigg(\dfrac{-\emph{p}\lambda^{\textrm{K}}_{Y}s\pi}{\sqrt{\big[m\cos(\theta_{{M}})\big]^2+s}}\Bigg).
\end{equation}
\textit{Proof}:  See Appendix B.\hfill $ \blacksquare $ \\

\section{Simulations and Discussions}
In this section, we evaluate the performance of cooperative NOMA at road intersections. 
In order to verify the accuracy of the theoretical results, Monte Carlo simulations are carried out by averaging over 10,000 realizations of the PPPs and fading parameters. In all figures, Monte Carlo simulations are presented by marks, and they match perfectly the theoretical results, which validates the correctness of our analysis. We set, without loss of generality,  $\lambda^{\textrm{LOS}}_X = \lambda^{\textrm{LOS}}_Y = \lambda^{\textrm{NLOS}}_X = \lambda^{\textrm{NLOS}}_Y =\lambda$. $S=(0,0), R=(50,0), D_1=(100,10), D_2=(100,-10)$, $\beta=9.5\times10^3$ \cite{bai2014analysis}, $\mu=1$. We set $\alpha_{\textrm{LOS}}=2$, $\alpha_{\textrm{NLOS}}=4$, $m_{\textrm{LOS}}=2$,  and  $m_{\textrm{NLOS}}=1$. Finally, we set $G_{max}=18$ dBi, $\eta=30$ GHz.
\begin{figure}[]
\centering
\includegraphics[height=8cm,width=9cm]{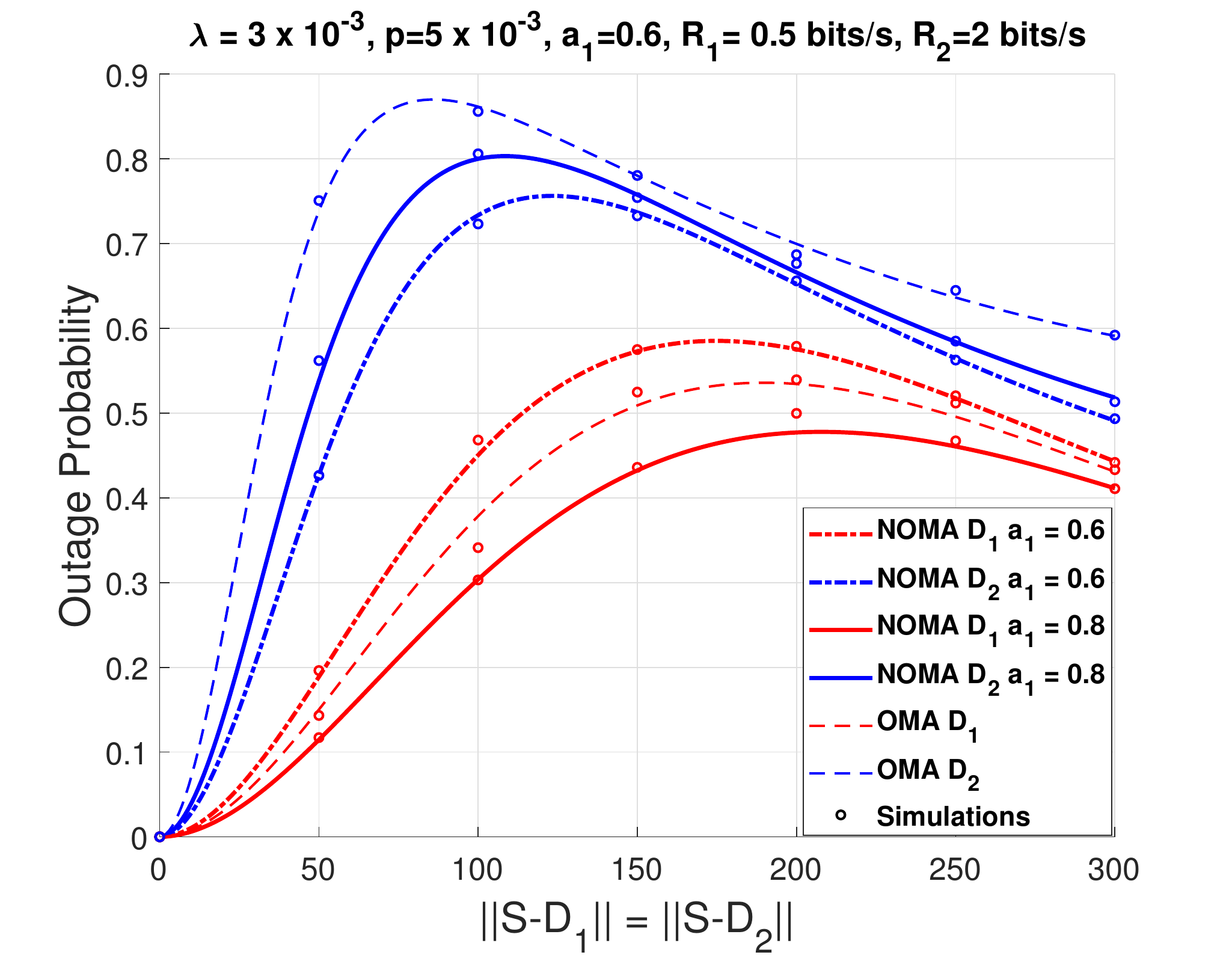}
\caption{Outage probability as a function of $\Vert S -D_1\Vert  =  \Vert S -D_2\Vert  $. The relay $R$ is always at mid distance between the source and the destination.}
\label{Fig2}
\end{figure}

Fig. \ref{Fig3} plots the outage probability as function of $\lambda$ considering cooperative NOMA, for LOS transmission, NLOS, and LOS/NLOS. We can see that LOS scenario has the highest outage probability. This is because, when the interference are in direct line of sight with the set $\{S,R,D_1,D_2\}$, the power of aggregate interference increases, hence reducing the SIR and increasing the outage. on the other hand, the NLOS scenario has the smallest outage, since the interference are in non  line of sight with the transmitting nodes. The model for this paper include a blockage model that includes both LOS and NLOS. Therefore, we wan see that the performance are between the LOS scenario and NLOS scenario, which are two extreme cases.

Fig.\ref{Fig2} plots the outage probability as a function of the distance between the source and the destinations.  Without loss of generality, we set $R$ at mid distance between $S$ and the two destinations $D_1$ and $D_2$. We can see that cooperative  NOMA outperforms cooperative OMA when $a_1=0.8$ for both $D_1$ and $D_2$. However, this is not the case for $a_1=0.6$, when NOMA outperforms OMA only for $D_2$. This is because when $a_1$ decreases, less power is allocated to $D_1$, hence it increases the outage probability. We can also see from Fig.\ref{Fig2} that the outage probability increases until 200 m for $D_1$ (100 m for $D_2$). This because, as the distance between the transmitting and the receiving nodes increases, the LOS probability decreases, and the NLOS probability increases, hence decreasing the outage probability. 
\begin{figure}[]
\centering
\includegraphics[height=8cm,width=9cm]{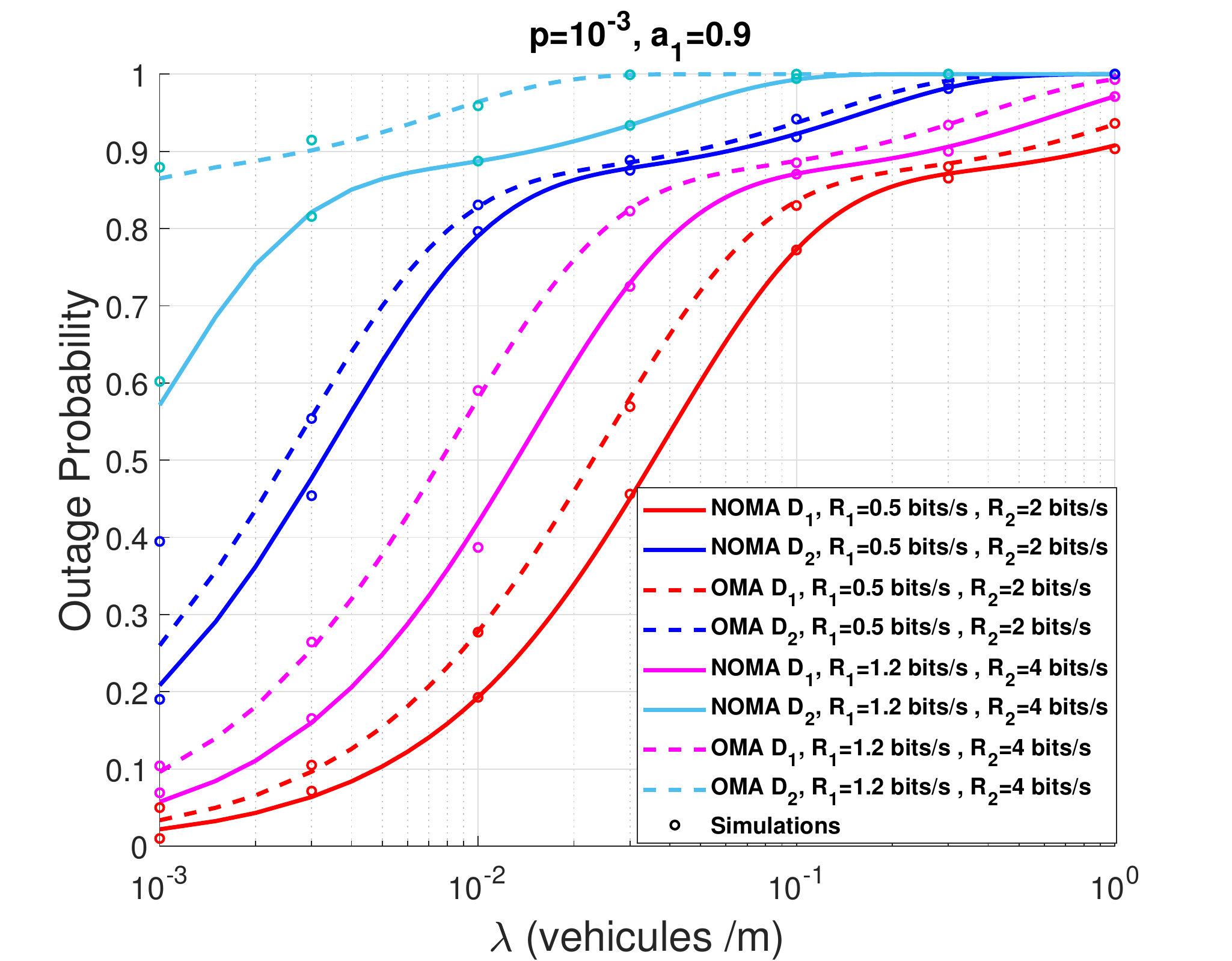}
\caption{Outage probability as a function of $\lambda$ considering cooperative NOMA and cooperative OMA.}
\label{Fig4}
\end{figure}
\begin{figure}[]
\centering
\includegraphics[height=8cm,width=9cm]{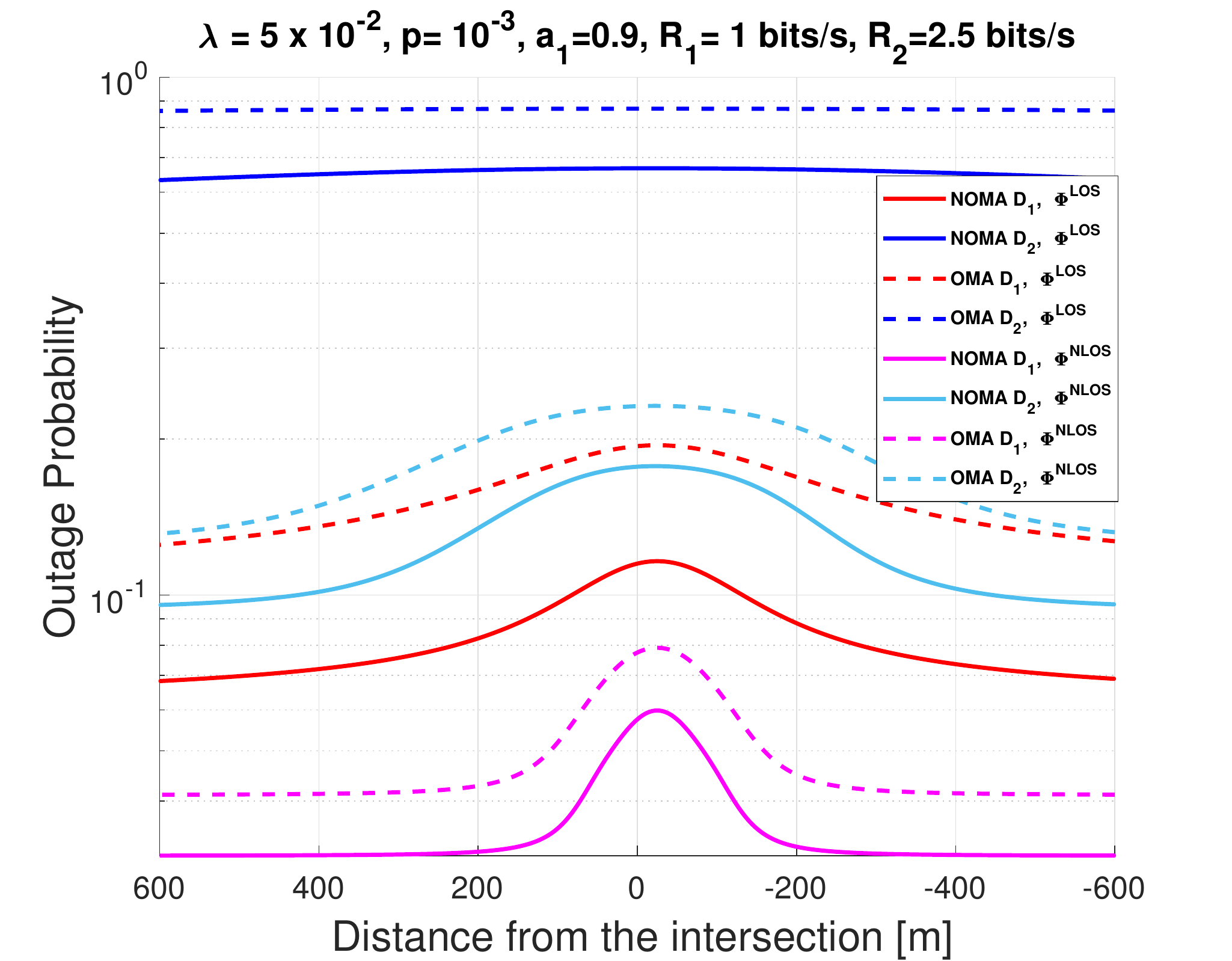}
\caption{Outage probability as a function of the distance form the intersection considering cooperative NOMA and cooperative OMA, for LOS scenarion and NLOS scenario.}
\label{Fig5}
\end{figure}

Fig.\ref{Fig4} plots the outage probability as a function of $\lambda$ considering cooperative  NOMA and cooperative  OMA for several values of data rates. 
We can see that NOMA outperforms OMA. We can also see that $D_1$ has a better performance than $D_2$. This is because $D_1$ has a smaller target data rate, since $D_1$ need to be served quickly (e.g., alert message). We can also see that, as the data rates increases ($R_1=1.2$bits/s and $R_2=4$bits/s), the gap of performance between NOMA and OMA increases. This is because, as the data rates increases, the decoding threshold of OMA increases dramatically ($\Theta_{\textrm{OMA}}=2^{4\mathcal{R}}-1$). The increase of the threshold becomes larger for $D_2$, since it has a higher data rate that $D_1$. 

Fig.\ref{Fig5} plots the outage probability of  the distance from the intersection considering cooperative NOMA and cooperative OMA, for LOS scenario and NLOS scenario. Without loss of generality, we set $R$ at mid distance between $S$ and the two destinations $D_1$ and $D_2$. We notice from Fig.\ref{Fig5}  that as nodes approach the intersection, the outage probability increases.  This because when the nodes are far from the intersection, only the interferes in the same road segment contribute to the aggregate interference, but as the node approach the intersection, both road segments contribute to the aggregate interference.
However, we can see that $D_2$ has a severe outage in LOS scenario compared to NLOS, and that the increases of the outage for $D_2$ in LOS, when the nodes move toward the intersection is negligible. This is because, in a LOS scenario,  the interferers from both road segment contributes the aggregate interference, whether the nodes are close or far away from the intersection. 
\section{Conclusion}

  In this paper, we studied cooperative NOMA for mmWave vehicular networks at intersection roads. The analysis was conducted using tools from stochastic geometry and was verified with Monte Carlo simulations.
  We derived closed form outage probability expressions for cooperative NOMA, and compared them with cooperative OMA. We showed that cooperative NOMA exhibited a significant improvement compared to cooperative OMA, especially for high data rates. However, data rates have to respect a given condition, if not, the performance of cooperative NOMA will decreases drastically. We also showed that as the nodes reach the intersection, the outage probability increased. Counter-intuitively, we showed that NLOS scenario has a better performance than LOS scenario.   
 
\appendices
\section{}\label{App_A}
To calculate $\mathbb{P}(\textit{O}_{(1)})$, we express it as a function of a success probability $\mathbb{P}(\textit{O}_{(1)}^C)$, where $\mathbb{P}(\textit{O}_{D_1}^C)$ is expressed as 
\begin{equation} \label{App.1}
\mathbb{P}(\textit{O}_{(1)})=1-\mathbb{P}(\textit{O}^C_{(1)}),
\end{equation}
The probability $\mathbb{P}(\textit{O}^C_{(1)})$ is expressed as
\begin{equation}\label{App.2}
\mathbb{P}(\textit{O}^C_{(1)})=1-\mathbb{P}(\textit{O}^C_{R_1} \cap \textit{O}^C_{D_1}) \\ 
=\mathbb{P}(\textit{O}^C_{R_1})\mathbb{P}(\textit{O}^C_{D_1}),
\end{equation}
where
\begin{equation}\label{App.3}
\textit{O}^C_{R_1}\triangleq \bigcup_{{\textrm{Z}}\in \{\textrm{LOS},\textrm{NLOS}\}} \big\{ \textrm{Z}_{SR}\cap (\textrm{SIR}^{(\alpha_{\textrm{Z}})}_{R_1} \geq \Theta_1) \big\}   
\end{equation}
\begin{equation}\label{App.4}
\textit{O}^C_{D_1}\triangleq \bigcup_{{\textrm{Z}}\in \{\textrm{LOS},\textrm{NLOS}\}} \big\{ \textrm{Z}_{RD_1}\cap (\textrm{SIR}^{(\alpha_{\textrm{Z}})}_{D_1} \geq \Theta_1) \big\} .  
\end{equation}
We calculate The probability $\mathbb{P}(\textit{O}^C_{R_1})$ as  
\begin{eqnarray}\label{App.8}
\mathbb{P}(\textit{O}^C_{R_1})&=&\sum^{}_{{\textrm{Z}}\in \{\textrm{LOS},\textrm{NLOS}\}} \mathbb{E}_{I_{X},I_Y}\Bigg[\mathbb{P}\Bigg\lbrace\ \textrm{Z}_{SR}\cap (\textrm{SIR}^{(\alpha_{\textrm{Z}})}_{R_1} \geq \Theta_1)\Bigg\rbrace\Bigg] \nonumber \\
&=&\sum^{}_{{\textrm{Z}}\in \{\textrm{LOS},\textrm{NLOS}\}}\mathbb{P}(\textrm{Z}_{SR})\:\:     \mathbb{E}_{I_{X},I_Y}\Bigg[\mathbb{P}\Bigg\lbrace\  \textrm{SIR}^{(\alpha_{\textrm{Z}})}_{R_1} \geq \Theta_1\Bigg\rbrace\Bigg] \nonumber \\
&=&\sum^{}_{{\textrm{Z}}\in \{\textrm{LOS},\textrm{NLOS}\}}\mathbb{P}(\textrm{Z}_{SR})\:\:     \mathbb{E}_{I_{X},I_Y}\Bigg[\mathbb{P}\Bigg\lbrace\ \frac{\vert h_{SR}\vert^{2}r_{SR}^{-\alpha_{\textrm{Z}}}\Upsilon a_1}{\vert h_{SR}\vert^{2}r_{SR}^{-\alpha_{{\textrm{Z}}}}\Upsilon a_2+I_{X_{R}}+I_{Y_{R}}} \ge \Theta_1\Bigg\rbrace\Bigg] \nonumber \\
&=&\sum^{}_{{\textrm{Z}}\in \{\textrm{LOS},\textrm{NLOS}\}}\mathbb{P}(\textrm{Z}_{SR})\:\:    \mathbb{E}_{I_{X},I_Y}\Bigg[\mathbb{P}\Bigg\lbrace\vert h_{SR}\vert^{2}r_{SR}^{-\alpha_{\textrm{Z}}} \Upsilon (a_1-\Theta_1 a_2)\ge\Theta_1\big[I_{X_{R}}+I_{Y_{R}}\big]\Bigg\rbrace\Bigg]. 
\end{eqnarray}
We can notice from (\ref{App.8}) that, when $\Theta_1 \ge a_1/ a_2$, the success probability $\mathbb{P}(\textit{O}^C_{R_1})$ is always zero, that is, $\mathbb{P}(\textit{O}_{R_1})=1$. Then, when $\Theta_1 < a_1/ a_2$, and after setting $\Psi_{1}= \Theta_1 /(a_1- \Theta_1 a_2)$, then
\begin{equation}\label{App.9}
\mathbb{P}(\textit{O}^C_{R_1}) =\sum^{}_{{\textrm{Z}}\in \{\textrm{LOS},\textrm{NLOS}\}}\mathbb{P}(\textrm{Z}_{SR})\:\:    \mathbb{E}_{I_{X},I_Y}\Bigg[\mathbb{P}\Bigg\lbrace\vert h_{SR}\vert^{2}\ge\frac{\Psi_{1}}{r_{SR}^{-\alpha_{\textrm{Z}} }\Upsilon  }\big[I_{X_{R}}+I_{Y_{R}}\big]\Bigg\rbrace\Bigg].\nonumber
\end{equation}
Since $\vert h_{SR}\vert^{2}$ follows a gamma distribution, its complementary cumulative distribution function (CCDF) is given by
\begin{equation}\label{App.10}
\bar{F}_{\vert h_{SR}\vert^{2}}(X)=\mathbb{P}(\vert h_{SR}\vert^{2}>X)=\frac{\Gamma(m_{\textrm{Z}},\frac{m_{\textrm{Z}}}{\mu }X)}{\Gamma(m_{\textrm{Z}})},
\end{equation}
hence   
\begin{eqnarray}\label{App.11}
\mathbb{P}(\textit{O}^C_{R_1})&=&\sum^{}_{{\textrm{Z}}\in \{\textrm{LOS},\textrm{NLOS}\}}\mathbb{P}(\textrm{Z}_{SR})\:\:   \mathbb{E}_{I_{X},I_Y}\Bigg[\frac{\Gamma\Big(m_{\textrm{Z}},\dfrac{m_{\textrm{Z}} \:\Psi_1}{\mu\: r_{SR}^{-\alpha_{{\textrm{Z}}}} \Upsilon } (I^{\textrm{LOS}}_{X_{R}}+I^{\textrm{LOS}}_{Y_{R}})\Big)}{\Gamma(m_{\textrm{Z}})}\Bigg]\nonumber\\  &\times & 
\mathbb{E}_{I_{X},I_Y}\Bigg[\frac{\Gamma\Big(m_{\textrm{Z}},\dfrac{m_{\textrm{Z}} \:\Psi_1}{\mu\: r_{SR}^{-\alpha_{{\textrm{Z}}}} \Upsilon } (I^{\textrm{NLOS}}_{X_{R}}+I^{\textrm{NLOS}}_{Y_{R}})\Big)}{\Gamma(m_{\textrm{Z}})}\Bigg]\nonumber \\
&=&\sum^{}_{{\textrm{Z}}\in \{\textrm{LOS},\textrm{NLOS}\}}\mathbb{P}(\textrm{Z}_{SR})\:\:   \prod^{}_{{\textrm{K}}\in \{\textrm{LOS},\textrm{NLOS}\}} \mathbb{E}_{I_{X},I_Y}\Bigg[\frac{\Gamma\Big(m_{\textrm{Z}},\dfrac{m_{\textrm{Z}} \:\Psi_1}{\mu\: r_{SR}^{-\alpha_{{\textrm{Z}}}} \Upsilon } (I^{\textrm{K}}_{X_{R}}+I^{\textrm{K}}_{Y_{R}})\Big)}{\Gamma(m_{\textrm{Z}})}\Bigg] 
\end{eqnarray}
The exponential sum function when $m_{\textrm{Z}}$ is an integer is defined as 
\begin{equation}\label{App.12}
e_{(m_{\textrm{Z}})}=\sum_{k=0}^{m_{\textrm{Z}}-1} \frac{(\frac{m_{\textrm{Z}}}{\mu }X)^k}{k!}=e^X \frac{\Gamma(m_{\textrm{Z}},\frac{m_{\textrm{Z}}}{\mu }X)}{\Gamma(m_{\textrm{Z}})},
\end{equation}
then
\begin{equation}\label{App.13}
\frac{\Gamma(m_{\textrm{Z}},\frac{m_{\textrm{Z}}}{\mu }X)}{\Gamma(m_{\textrm{Z}})}=e^{-\frac{m_{\textrm{Z}}}{\mu}X}\sum_{k=0}^{m_{\textrm{Z}}-1}\frac{1}{k!}{\big(\frac{m_{\textrm{Z}}\:X}{\mu}\big)}^k.
\end{equation}
We denote the expectation in equation (\ref{App.11}) by $\mathcal{E}(I_{X},I_Y)$, then $\mathcal{E}(I_{X},I_Y)$ equals
\begin{eqnarray}\label{App.14}
\mathcal{E}(I_{X},I_Y)&=&\mathbb{E}_{I_{X},I_Y}\Bigg[\exp\Big(-\dfrac{m_{\textrm{Z}} \:\Psi_1}{\mu\: r_{SR}^{-\alpha_{\textrm{Z}}}\Upsilon } (I^{\textrm{K}}_{X_{R}}+I^{\textrm{K}}_{Y_{R}})\Big)\times  \sum_{k=0}^{m_{\textrm{Z}}-1}\frac{1}{k!}\Big(\dfrac{m_{\textrm{Z}}\:\Psi_1}{\mu \:r_{SR}^{-\alpha_{\textrm{Z}}}\Upsilon } (I^{\textrm{K}}_{X_{R}}+I^{\textrm{K}}_{Y_{R}})\Big)^k\Bigg]\nonumber\\
&=&\sum_{k=0}^{m-1}\frac{1}{k!}\:\Big(\dfrac{m_{\textrm{Z}}\:\Psi_1}{\mu \:r_{SR}^{-\alpha_{\textrm{Z}}}\Upsilon }\Big)^k \mathbb{E}_{I_{X},I_Y}\Bigg[\exp\Big(-\dfrac{m_{\textrm{Z}} \:\Psi_1}{\mu\: r_{SR}^{-\alpha_{\textrm{Z}}}\Upsilon } \big(I^{\textrm{K}}_{X_{R}}+I^{\textrm{K}}_{Y_{R}}\big)\Big)
\big(I^{\textrm{K}}_{X_{R}}+I^{\textrm{K}}_{Y_{R}}\big)^k\Bigg].
\end{eqnarray}
Applying the binomial theorem in (\ref{App.14}), we get
\begin{eqnarray}\label{App.15}
\mathcal{E}(I_{X},I_Y)&=&\sum_{k=0}^{m_{\textrm{Z}}-1}\frac{1}{k!}\:\Big(\dfrac{m_{\textrm{Z}}\:\Psi_1}{\mu \:r_{SR}^{-\alpha_{\textrm{Z}}}\Upsilon }\Big)^k \mathbb{E}_{I_{X},I_Y}\Bigg[\exp\Big(-\dfrac{m_{\textrm{Z}} \:\Psi_1}{\mu\: r_{SR}^{-\alpha_{\textrm{Z}}}\Upsilon } \big[I^{\textrm{K}}_{X_{R}}+I^{\textrm{K}}_{Y_{R}}\big]\Big)\sum_{n=0}^{k} \binom{k}{n}{(I^{\textrm{K}}_{X_{R}})}^{k-n}\:{(I^{\textrm{K}}_{Y_{R}})}^n\Bigg]\nonumber\\
&=&\sum_{k=0}^{m_{\textrm{Z}}-1}\frac{1}{k!}\:\Omega^k \mathbb{E}_{I_{X},I_Y}\Bigg[\exp\Big(-\Omega \big[I^{\textrm{K}}_{X_{R}}+I^{\textrm{K}}_{Y_{R}}\big]\Big)\sum_{n=0}^{k} \binom{k}{n}{(I^{\textrm{K}}_{X_{R}})}^{k-n}\:({I^{\textrm{K}}_{Y_{R}}})^n\Bigg],
\end{eqnarray}
where $\Omega=\dfrac{m_{\textrm{Z}}\:\Psi_1}{\mu \:r_{SR}^{-\alpha_{\textrm{Z}}} \Upsilon}$. To calculate the expectation in (\ref{App.15}) we process as follows

\begin{multline}\label{App.16}
\mathbb{E}_{I_{X},I_Y}\Bigg[e^{-\Omega\:I^{\textrm{K}}_{X_R}}\:e^{-\Omega\:I^{\textrm{K}}_{Y_R}}\sum_{n=0}^{k} \binom{k}{n}{(I^{\textrm{K}}_{X_{R}})}^{k-n}\:{(I^{\textrm{K}}_{Y_{R}})}^n\Bigg]= \sum_{n=0}^{k}\binom{k}{n}\mathbb{E}_{I_{X},I_Y}\Bigg[e^{-\Omega\:I^{\textrm{K}}_{X_{R}}}\:e^{-\Omega\:I^{\textrm{K}}_{Y_R}}{(I^{\textrm{K}}_{X_{R}})}^{k-n}\:{(I^{\textrm{K}}_{Y_{R}})}^n\Bigg]\nonumber
\end{multline}
\begin{eqnarray}\label{App.17}
&=&\sum_{n=0}^{k}\binom{k}{n}\mathbb{E}_{I_{X}}\Bigg[e^{-\Omega\:I^{\textrm{K}}_{X_R}}{(I^{\textrm{K}}_{X_R})}^{k-n}\Bigg]\mathbb{E}_{I^{\textrm{K}}_{Y_R}}\Bigg[e^{-\Omega\:I^{\textrm{K}}_{Y_R}}\:{(I^{\textrm{K}}_{Y_R})}^n\Bigg]\nonumber \\
&\overset{(a)}{=}&\sum_{n=0}^{k}\binom{k}{n}(-1)^{k-n}\frac{\textrm{d}^{k-n} \mathcal{L}_{I^{\textrm{K}}_{X_R}}(\Omega)}{\textrm{d}^{k-n} \Omega} (-1)^{n}\frac{\textrm{d}^{n} \mathcal{L}_{I^{\textrm{K}}_{Y_R}}(\Omega)}{\textrm{d}^{n} \Omega}\nonumber \\
&=&(-1)^{k}\sum_{n=0}^{k}\binom{k}{n}\frac{\textrm{d}^{k-n} \mathcal{L}_{I^{\textrm{K}}_{X_R}}(\Omega)}{\textrm{d}^{k-n} \Omega} \frac{\textrm{d}^{n} \mathcal{L}_{I^{\textrm{K}}_{Y_R}}(\Omega)}{\textrm{d}^{n} \Omega}. 
\end{eqnarray}
where (a) stems form the following property 
\begin{equation}\label{App.20}
\mathbb{E}_{I}\Big[e^{-\Omega I}{I}^{N}\Big]=(-1)^N \frac{\textrm{d}^{N}\mathbb{E}_{I}\Big[e^{-\Omega\:I}{I}^{N}\Big]}{\textrm{d}^{N} \Omega}=(-1)^N\frac{\textrm{d}^{N} \mathcal{L}_{I}(\Omega)}{\textrm{d}^{N} \Omega},
\end{equation}
Finally, the expectation becomes
\begin{equation}\label{App.21}
\sum_{k=0}^{m_{\textrm{Z}}-1}\frac{1}{k!}\:\big(-\dfrac{m_{\textrm{Z}}\:\Psi_1}{\mu \:r_{SR}^{-\alpha_{\textrm{Z}}}\Upsilon }\big)^k \sum_{n=0}^{k}\binom{k}{n}\frac{\textrm{d}^{k-n} \mathcal{L}_{I^{\textrm{K}}_{X_R}}\big(\dfrac{m_{\textrm{Z}}\:\Psi_1}{\mu \:r_{SR}^{-\alpha_{\textrm{Z}}}\Upsilon }\big)}{\textrm{d}^{k-n} \big(\dfrac{m_{\textrm{Z}}\:\Psi_1}{\mu \:r_{SR}^{-\alpha_{\textrm{Z}}} \Upsilon}\big)} \frac{\textrm{d}^{n} \mathcal{L}_{I^{\textrm{K}}_{Y_R}}\big(\dfrac{m_{\textrm{Z}}\:\Psi_1}{\mu \:r_{SR}^{-\alpha_{\textrm{Z}}}\Upsilon }\big)}{\textrm{d}^{n} \big(\dfrac{m_{\textrm{Z}}\:\Psi_1}{\mu \:r_{SR}^{-\alpha_{\textrm{Z}}}\Upsilon }\big)}.
\end{equation}
Then plugging (\ref{App.21}) in (\ref{App.11}) yields
\begin{multline}\label{App.22}
\mathbb{P}(\textit{O}_{R_1}^C)= \sum_{\textrm{Z}\in\{\textrm{LOS},\textrm{NLOS}\}}^{} \mathbb{P}(\textrm{Z}_{SR}) \times \\
\prod_{\textrm{K}\in\{\textrm{LOS},\textrm{NLOS}\}}\sum_{k=0}^{m_L-1}\frac{1}{k!}\:\big(-\dfrac{m_\textrm{Z}\:\Psi_{1}}{\mu \:r_{SR}^{-\alpha_\textrm{Z}} \Upsilon}\big)^k \sum_{n=0}^{k}\binom{k}{n}\frac{\textrm{d}^{k-n} \mathcal{L}_{I^{\textrm{K}}_{X_R}}\big(\dfrac{m_\textrm{Z}\:\Psi_{1}}{\mu \:r_{SR}^{-\alpha_\textrm{Z}}\Upsilon }\big)}{\textrm{d}^{k-n} \big(\dfrac{m_\textrm{Z}\:\Psi_{1}}{\mu \:r_{SR}^{-\alpha_\textrm{Z}}\Upsilon }\big)} \frac{\textrm{d}^{n} \mathcal{L}_{I^{\textrm{K}}_{Y_R}}\big(\dfrac{m_\textrm{Z}\:\Psi_{1}}{\mu \:r_{SR}^{-\alpha_\textrm{Z}}\Upsilon }\big)}{\textrm{d}^{n} \big(\dfrac{m_\textrm{Z}\:\Psi_{1}}{\mu \:r_{SR}^{-\alpha_\textrm{Z}} \Upsilon}\big)}\\
\end{multline}

The expression of $\textrm{d}^{k-n} \mathcal{L}_{I^{\textrm{K}}_{X}}(s)/\textrm{d}^{k-n} (s)$ and $\textrm{d}^{n} \mathcal{L}_{I^{\textrm{K}}_{Y}}(s)/\textrm{d}^{n} (s)$ are given by (\ref{eq.40}) and (\ref{eq.41}).
The probability $\mathbb{P}(\textit{O}^C_{D_1})$ can be calculated following the same steps above.

 In the same way we express $\mathbb{P}(\textit{O}_{(2)})$ as a function of a success probability $\mathbb{P}(\textit{O}_{(2)}^C)$, where $\mathbb{P}(\textit{O}_{(2)}^C)$ is given by
 \begin{equation}\label{App.23}
\mathbb{P}(\textit{O}_{(2)})=1-\mathbb{P}(\textit{O}^C_{(2)}).
\end{equation}
The probability $\mathbb{P}(\textit{O}^C_{(2)})$ is expressed as
\begin{equation} \label{App.24}
\mathbb{P}(\textit{O}^C_{(2)})=1-\mathbb{P}(\textit{O}^C_{R_2} \cap \textit{O}^C_{D_2}) \\ 
=\mathbb{P}(\textit{O}^C_{R_2})\mathbb{P}(\textit{O}^C_{D_2}),
\end{equation}
where
\begin{equation}\label{App.25}
\textit{O}^C_{R_2}\triangleq \bigcup_{{\textrm{Z}}\in \{\textrm{LOS},\textrm{NLOS}\}}\bigcap_{i=1}^{2} \big\{ \textrm{Z}_{SR} \cap (\textrm{SIR}^{(\alpha_{\textrm{Z}})}_{R_i} \geq \Theta_i) \big\}   
\end{equation}
\begin{equation}\label{App.26}
\textit{O}^C_{D_2}\triangleq \bigcup_{{\textrm{Z}}\in \{\textrm{LOS},\textrm{NLOS}\}}\bigcap_{i=1}^{2} \big\{ \textrm{Z}_{RD_2} \cap (\textrm{SIR}^{(\alpha_{\textrm{Z}})}_{D_{2-i}}< \Theta_i) \big\} .  
\end{equation}
To calculate $\mathbb{P}(\textit{O}^C_{R_2})$ we proceed as follows
\begin{eqnarray} \label{App.27}
\mathbb{P}(\textit{O}^C_{R_2})&=&\sum^{}_{{\textrm{Z}}\in \{\textrm{LOS},\textrm{NLOS}\}} \mathbb{E}_{I_{X},I_Y}\Bigg[\mathbb{P}\Bigg\lbrace\ \bigcap_{i=1}^{2} \big\{ \textrm{Z}_{SR} \cap (\textrm{SIR}^{(\alpha_{\textrm{Z}})}_{R_i} \geq \Theta_i)\Bigg\rbrace\Bigg] \nonumber \\
&=&\sum^{}_{{\textrm{Z}}\in \{\textrm{LOS},\textrm{NLOS}\}}\mathbb{P}(\textrm{Z}_{SR})\:\:      \mathbb{E}_{I_{X},I_Y}\Bigg[\mathbb{P}\Bigg\lbrace\ \bigcap_{i=1}^{2}  \textrm{SIR}^{(\alpha_{\textrm{Z}})}_{R_i} \geq \Theta_i\Bigg\rbrace\Bigg] \nonumber \\
&=&\sum^{}_{{\textrm{Z}}\in \{\textrm{LOS},\textrm{NLOS}\}}\mathbb{P}(\textrm{Z}_{SR})\:\:      \mathbb{E}_{I_{X},I_Y}\Bigg[\mathbb{P}\Bigg\lbrace\  \textrm{SIR}^{(\alpha_{\textrm{Z}})}_{R_{1}} \geq \Theta_1 \cap \textrm{SIR}^{(\alpha_{\textrm{Z}})}_{R_2} \geq \Theta_2\Bigg\rbrace\Bigg]. \nonumber \\
\end{eqnarray}

Following the same steps as for $\mathbb{P}(\textit{O}_{R_1}^C)$, we get
\begin{equation} \label{App.30}
\mathbb{P}(\textit{O}_{R_2}^C) =\mathbb{E}_{I_{X},I_Y}\Bigg[\mathbb{P}\Bigg\lbrace\frac{\vert h_{SR}\vert^{2}r_{SR}^{-\alpha_{\textrm{Z}}} \Upsilon a_1}{\vert h_{SR}\vert^{2}r_{SR}^{-\alpha_{\textrm{Z}}} \Upsilon a_2+I_{X_{R}}+I_{Y_{R}}} \ge \Theta_1, \frac{\vert h_{SR}\vert^{2}r_{SR}^{-\alpha_{\textrm{Z}}} \Upsilon a_2}{I_{X_{R}}+I_{Y_{R}}} \ge \Theta_2\Bigg\rbrace\Bigg].\nonumber
\end{equation}
When $\Theta_1 > a_1/ a_2$, then $\mathbb{P}(\textit{O}_{R_2})=1$, 
otherwise we continue the derivation
We set $\Psi_{2}= \Theta_2/a_2$, then
\begin{eqnarray} \label{App.31}
\mathbb{P}(\textit{O}_{R_2}^C)&=&\mathbb{E}_{I_{X},I_Y}\Bigg[\mathbb{P}\Bigg\lbrace\vert h_{SR}\vert^{2}\ge\frac{\Psi_{1}}{r_{SR}^{-\alpha_{\textrm{Z}}} \Upsilon }\big[I_{X_{R}}+I_{Y_{R}}\big], \vert h_{SR}\vert^{2}\ge\frac{\Psi_{2}}{r_{SR}^{-\alpha_{\textrm{Z}}} \Upsilon }\big[I_{X_{R}}+I_{Y_{R}}\big]\Bigg\rbrace\Bigg]\nonumber \\
&=&\mathbb{E}_{I_{X},I_Y}\Bigg[\mathbb{P}\Bigg\lbrace\vert h_{SR}\vert^{2}\ge\frac{\mathrm{max}(\Psi_{1},\Psi_{2})}{r_{SR}^{-\alpha_{\textrm{Z}}} \Upsilon }\big[I_{X_{R}}+I_{Y_{R}}\big]\Bigg\rbrace\Bigg].\nonumber
\end{eqnarray}
Following the same steps above, $\mathbb{P}(\textit{O}_{R_2}^C)$ equals
\begin{multline} \label{App.33}
\mathbb{P}(\textit{O}_{R_2}^C)= \sum_{\textrm{Z}\in\{\textrm{LOS},\textrm{NLOS}\}}^{} \mathbb{P}(\textrm{Z}_{SR}) \times \\
\prod_{\textrm{K}\in\{\textrm{LOS},\textrm{NLOS}\}}\sum_{k=0}^{m_L-1}\frac{1}{k!}\:\big(-\dfrac{m_\textrm{Z}\:\Psi_{\mathrm{max}}}{\mu \:r_{SR}^{-\alpha_\textrm{Z}} \Upsilon}\big)^k \sum_{n=0}^{k}\binom{k}{n}\frac{\textrm{d}^{k-n} \mathcal{L}_{I^{\textrm{K}}_{X_R}}\big(\dfrac{m_\textrm{Z}\:\Psi_{\mathrm{max}}}{\mu \:r_{SR}^{-\alpha_\textrm{Z}}\Upsilon }\big)}{\textrm{d}^{k-n} \big(\dfrac{m_\textrm{Z}\:\Psi_{\mathrm{max}}}{\mu \:r_{SR}^{-\alpha_\textrm{Z}}\Upsilon }\big)} \frac{\textrm{d}^{n} \mathcal{L}_{I^{\textrm{K}}_{Y_R}}\big(\dfrac{m_\textrm{Z}\:\Psi_{\mathrm{max}}}{\mu \:r_{SR}^{-\alpha_\textrm{Z}}\Upsilon }\big)}{\textrm{d}^{n} \big(\dfrac{m_\textrm{Z}\:\Psi_{\mathrm{max}}}{\mu \:r_{SR}^{-\alpha_\textrm{Z}} \Upsilon}\big)}\\
\end{multline}
where $\Psi_{\mathrm{max}}=\mathrm{max}(\Psi_1,\Psi_2)$. The probability $\mathbb{P}(\textit{O}_{D_2}^C)$ can be calculated following the same steps above.

\section{}\label{App_B}
The Laplace transform of the interference originating from the X road at $M$ is expressed as 
\begin{eqnarray}\label{eq:64}
\mathcal{L}_{{I^{\textrm{K}}_{X_{M}}}}(s) 
&=&\mathbb{E}\Bigg[{\exp\big(-sI^{\textrm{K}}_{X_{M}} \big)}\Bigg]\nonumber \\
&=&\mathbb{E}\Bigg[{\exp\Bigg(-\sum_{x\in\Phi^{\textrm{K}}_{X_{M}}}s\vert h_{{M}x}\vert^2 r_{{M}x}^{-\alpha_{\textrm{K}}}  \Bigg)}\Bigg]\nonumber \\
&=& \mathbb{E}\Bigg[\prod_{x\in\Phi^{\textrm{K}}_{X_{M}}} \exp\Bigg(-s\vert h_{{M}x}\vert^2 r_{{M}x}^{-\alpha_{\textrm{K}}}\Bigg)\Bigg]\nonumber \\
&\overset{(a)}{=}&\mathbb{E}\Bigg[\prod_{x\in\Phi^{\textrm{K}}_{X_{M}}}\mathbb{E}_{\vert  h_{{M}x}\vert^2, p}\Bigg\lbrace \exp\Bigg(-s\vert h_{{M}x}\vert^2r_{Mx}^{-\alpha_{\textrm{K}}}\Bigg)\Bigg\rbrace\Bigg]\nonumber \\
&\overset{(b)}{=}&\mathbb{E}\Bigg[\prod_{x\in\Phi^{\textrm{K}}_{X_{M}}}\dfrac{p}{1+s r_{{M}x}^{-\alpha_{\textrm{K}}}}+1-p\Bigg]\nonumber \\
&\overset{(c)}{=}&\exp\Bigg(-\lambda^{\textrm{K}}_{X}\displaystyle\int_{\mathbb{R}}\Bigg[1-\bigg(\dfrac{p}{1+sr_{{M}x}^{-\alpha_{\textrm{K}}}}+1-p\bigg)\Bigg]\textrm{d}x\Bigg)\nonumber \\
&=&\exp\Bigg(-p\lambda^{\textrm{K}}_{X}\displaystyle\int_{\mathbb{R}}\dfrac{1}{1+1/sr_{{M}x}^{-\alpha_{\textrm{K}}}}\textrm{d}x\Bigg)\\
&=&\exp\Bigg(-p\lambda^{\textrm{K}}_{X}\displaystyle\int_{\mathbb{R}}\dfrac{1}{1+r_{{M}x}^{\alpha_{\textrm{K}}}/s}\textrm{d}x\Bigg), 
\end{eqnarray}

where (a) follows from the independence of the fading coefficients; (b) follows from performing the expectation over $|h_{{M}x}|^2$ which follows an exponential distribution with unit mean, and performing the expectation over the set of interferes; (c) follows from the probability generating functional (PGFL) of a PPP. The expression of $\mathcal{L}_{{I^{\textrm{K}}_{Y_{M}}}}(s)$ can be acquired by following the same steps. 
The Laplace transform of the interference originating from the X road at the received node denoted  $M$, is expressed as

\begin{equation}\label{eq:33}
\mathcal{L}_{I^{\textrm{K}}_{X_{M}}}(s)=\exp\Bigg(-\emph{p}\lambda^{\textrm{K}}_{X}\int_\mathbb{R}\dfrac{1}{1+\Vert \textit{x}-{M} \Vert^{\alpha_{\textrm{K}}}/s}\textrm{d}x\Bigg),
\end{equation}
where 
\begin{equation}\label{eq:34}
\Vert \textit{x}-{M} \Vert=\sqrt{\big[m\sin(\theta_{{M}})\big]^2+\big[x-m \cos(\theta_{M}) \big]^2 }.
\end{equation}

The Laplace transform of the interference originating from the Y road at $M$ is given by
 
\begin{equation}\label{eq:35}
\mathcal{L}_{I^{\textrm{K}}_{Y_{M}}}(s)=\exp\Bigg(-\emph{p}\lambda^{\textrm{K}}_{Y}\int_\mathbb{R}\dfrac{1}{1+\Vert \textit{y}-{M} \Vert^{\alpha_{\textrm{K}}}/s}\textrm{d}y\Bigg),
\end{equation}
where
\begin{equation}\label{eq:36}
\Vert \textit{y}-{M} \Vert=\sqrt{\big[m\cos(\theta_{{M}})\big]^2+\big[y-m \sin(\theta_{M}) \big]^2 },
\end{equation}
where $\theta_{M} $ is the angle between the node ${M}$ and the X road.

In order to calculate the Laplace transform of interference originated from the X road at the node $M$, we have to calculate the integral in (\ref{eq:33}). We calculate the integral in (\ref{eq:33}) for $\alpha_{\textrm{K}}=2$. Let us take $m_{x}=m \cos(\theta_{M})$, and $m_{y}=m \sin(\theta_{M}$), then (\ref{eq:33}) becomes
\begin{eqnarray}\label{eq:65}
\mathcal{L}_{I^{\textrm{K}}_{X_{M}}}(s)&=&\exp\Bigg(-\emph{p}\lambda^{\textrm{K}}_{X}\int_\mathbb{R}\dfrac{1}{1+m_{y}^2+(x-m_{x})^2/s }\textrm{d}x\Bigg),\nonumber\\
&=&\exp\Bigg(-\emph{p}\lambda^{\textrm{K}}_{X}s\int_\mathbb{R}\dfrac{1}{s+m_{y}^2+(x-m_{x})^2 }\textrm{d}x\Bigg),
\end{eqnarray}
and the integral inside the exponential in (\ref{eq:65}) equals
\begin{equation}\label{eq:66}
\int_\mathbb{R}\dfrac{1}{s+m_{y}^2+(x-m_{x})^2}\textrm{d}x=\dfrac{\pi}{\sqrt{m_{y}^2+s}}.
\end{equation}
Then, plugging (\ref{eq:66}) into (\ref{eq:65}), and substituting $m_{y}$ by $m\sin(\theta_{{M}})$  we obtain
\begin{equation}\label{eq:67}
\mathcal{L}_{I^{\textrm{K}}_{X_M}}(s)=\exp\Bigg(-\dfrac{\emph{p}\lambda^{\textrm{K}}_{X}s\,\pi}{\sqrt{m^2\sin(\theta_{{M}})^2+s}}\Bigg).
\end{equation}
Following the same steps above, and without details for the derivation with respect to $s$, we obtain 
\begin{equation}\label{eq:68}
\mathcal{L}_{I^{\textrm{K}}_{Y_M}}(s)=\exp\Bigg(-\dfrac{\emph{p}\lambda^{\textrm{K}}_{Y}s\,\pi}{\sqrt{m^2\cos(\theta_{{M}})^2+s}}\Bigg).
\end{equation}
Then, when compute the derivative of (\ref{eq:67}) and (\ref{eq:68}), we obtain

\begin{multline}\label{eq.40}  
\frac{\textrm{d}^{k-n} \mathcal{L}_{I^{\textrm{K}}_{X_{M}}}\big(s\big)}{\textrm{d}^{k-n} s}= \Bigg[-\frac{\emph{p}\lambda^{\textrm{K}}_{X}\pi}{\sqrt{t^2\sin(\theta_{{M}})^2+s}}+\frac{1}{2}\frac{\emph{p}\lambda^{\textrm{K}}_{X}\pi s}{(m^2\sin(\theta_{{M}})^2+s)^{3/2}}\Bigg]^{k-n}\\
\times \exp\Bigg(-\frac{\emph{p}\lambda^{\textrm{K}}_{X}\pi s}{\sqrt{m^2\sin(\theta_{{M}})^2+s}}\Bigg).
\end{multline}
\begin{multline}\label{eq.41} 
  \frac{\textrm{d}^{n} \mathcal{L}_{I^{\textrm{K}}_{Y_{M}}}\big(s\big)}{\textrm{d}^{n} s}=\Bigg[-\frac{\emph{p}\lambda^{\textrm{K}}_{Y}\pi}{\sqrt{m^2\cos(\theta_{{M}})^2+s}}+\frac{1}{2}\frac{\emph{p}\lambda^{\textrm{K}}_{Y}\pi s}{(m^2\cos(\theta_{{M}})^2+s)^{3/2}}\Bigg]^n \\
\times \exp\Bigg(-\frac{\emph{p}\lambda^{\textrm{K}}_{Y}\pi s}{\sqrt{m^2\cos(\theta_{{M}})^2+s}}\Bigg).
\end{multline} 

\bibliographystyle{ieeetr}
\bibliography{bibnoma}

\end{document}